\documentclass[final,1p,times]{elsarticle}

\usepackage{amssymb}
\journal{Computer Physics Communications}

\begin{document}

\begin{frontmatter}

\author[auth1]{N.~Dubray}
\address[auth1]{CEA, DAM, DIF, F-91297 Arpajon, France}
\ead{noel.dubray@cea.fr}

\author[auth2]{D.~Regnier}
\address[auth2]{CEA, DEN, DER, F-13108 Saint Paul lez Durance, France}

\title{Numerical search of discontinuities in self-consistent potential energy surfaces}

\begin{abstract}
Potential energy surfaces calculated with self-consistent mean-field methods are a very powerful tool, since their solutions are, in theory, global minima of the non-constrained subspace. However, this minimization leads to an incertitude concerning the saddle points, that can sometimes be no more saddle points in bigger constrained subspaces ({\it fake} saddle points), or can be missing on a trajectory ({\it missing} saddle points). These phenomena are the consequences of discontinuities of the self-consistent potential energy surfaces (SPES). These discontinuities may have important consequences, since they can for example hide the real height of an energy barrier, and avoid any use of a SPES for further dynamical calculations, barrier penetrability estimations, or trajectory predictions. Discontinuities are not related to the quality of the production of a SPES, since even a perfectly converged SPES with an ideally fine mesh can be discontinuous. In this paper we explain what are the discontinuities, their consequences, and their origins. We then propose a numerical method to detect and identify discontinuities on a given SPES, and finally we discuss what are the best ways to transform a discontinuous SPES into a continuous one.
\end{abstract}


\end{frontmatter}

\section*{Introduction}

Potential energy surfaces (PES) are a widely used tool to describe physical and chemical systems, among others. For example, they are used in DNA chemistry~\cite{Elsawy:2005}, materials chemistry~\cite{Buehler:2000}, chemical physics~\cite{Roy:2007}, astrophysics~\cite{Schwartz:2010}, etc\ldots
Numerous methods exist to extract from a PES local minima, saddle points, and least energy paths between local minima~\cite{Henkelman:2000,Olsen:2004}. To produce a PES, a method giving the energy of a system as a function of a given number of constrained variables is needed. In this paper, we will separate these methods into two main classes: self-consistent and non-self-consistent methods. Surfaces associated with these methods will be called Self-consistent Potential Energy Surfaces (SPES) and Non Self-consistent Energy Surfaces (NSPES), respectively. In nuclear physics, SPES can be obtained for example by constrained Hartree-Fock methods and extensions~\cite{Warda:2002,Rodriguez:2010,Hinohara:2009,Bonneau:2006}, by constrained Relativistic Mean Field method~\cite{Guo:2010}, etc\ldots. NSPES can be produced by several methods, ranging from the historical liquid drop model~\cite{Meitner:1939,Bohr:1939} to the well-known macroscopic-microscopic model, using parametrization of the nuclear mean-field deformation~\cite{Moller:2001,Pyatkov:1997}.

The main difference between these two classes of methods is the way they deal with the non-constrained degrees of freedom of the system. Non self-consistent methods neglect their influence (all non-constrained degrees of freedom take a fixed value), while self-consistent methods perform an automatic minimization of the energy of the system in the non-constrained subspace. The presence or the absence of this automatic minimization leads to one specific problem for each class of methods:
\begin{itemize}
 \item {\bf NSPES problem}: every point of a given NSPES may not be a minimum of the same NSPES with an additional dimension.
 \item {\bf SPES problem}: a saddle point on a trajectory from a SPES is not automatically visible on the same trajectory from the same SPES restricted to a smaller dimension ({\it missing} saddle point), and a saddle point from a SPES is not automatically a saddle point on the same SPES with an additional dimension ({\it fake} saddle point).
\end{itemize}

These two problems have important consequences. For the NSPES problem, one would expect that enriching a NSPES with an additional constraint while conserving a given set of symmetries should improve the quality of the description, instead of invaliding it, but this is not always the case. For example, quantities calculated on a $N$-dimensional NSPES like minima, saddle points, or least energy paths between local minima {\it may differ dramatically} from quantities calculated on a $(N+1)$-dimensional NSPES, with the first $N$ constraints and conserved symmetries being the same for both NSPES.
Let us imagine an hypothetical non-constrained HFB minimum $|\psi\rangle$ with $\langle \psi|\hat{Q}_{70}|\psi\rangle=$~37~b$^{7/2}$. Finding this minimum with non-self-consistent methods will require to explore the deformation degree of freedom corresponding to $\hat{Q}_{70}$, without knowing beforehand that this degree of freedom has a role to play. If this degree of freedom is not taken into account, the resulting minimum can be anything but the searched minimum, and may differ dramatically from it.

Concerning the SPES problem, the obtained saddle points can be {\it fake} or {\it missing}. This last problem is well known (it corresponds to a {\it tipping point} in {\it Catastrophe Theory}~\cite{Gilmore:1981}), and is used sometimes to contest -with good reason- the quality of the results obtained by self-consistent or minimization methods~\cite{Myers:1996}. For example, in~\cite{Moller:2009}, the authors present a schematic case leading to the calculation of a {\it fake} saddle point by a minimization method, and conclude rather quickly that the doubt concerning the reality of saddle points found with self-consistent methods invalidates any production of a SPES by a minimization method: ``[\ldots] in Hartree-Fock-Bogoliubov (HFB) calculations with multiple constraints, saddle-point shapes and energies frequently cannot be determined accurately and are subject to errors of fairly random magnitude".

To our knowledge, there is no numerical or analytical method to fix the NSPES problem. Any result extracted from a NSPES may be invalidated by a NSPES of higher dimension. By comparison, there exists a simple numerical method to fix the SPES problem, that we will present in this paper.

A saddle point that is no more a saddle point in a bigger constrained subspace and a saddle point which vanishes while reducing the constrained subspace are two of the possible consequences of a unique phenomenon that we call a discontinuity of the SPES. A continuous SPES has no {\it fake} or {\it missing} saddle point. In this paper we first show such discontinuities, explain some of their consequences, and explain their origins. We then propose a method to quantitatively estimate if a SPES is continuous or not, and to identify the discontinuities if present. Finally we discuss what are the best ways to get a continuous SPES from a discontinuous one.

\section{Discontinuities}

\subsection{Model used}

Every calculation presented in this paper has been obtained by a two-center basis Hartree-Fock-Bogoliubov model \cite{Ring:1980} with Gogny D1S nucleon-nucleon effective interaction~\cite{Decharge:1980,Berger:1991}, based on the minimization principle of the energy functional, namely
\begin{equation}
\delta\langle \psi|\hat{H}-\lambda_N\hat{N}-\lambda_Z\hat{Z}-\sum_l\lambda_l\hat{Q}_{l0}|\psi\rangle=0\label{HFB1},
\end{equation}
where $\hat{H}$ is the nuclear microscopic Hamiltonian, $\hat{Q}_{l0}$ is a mass multipole operator, and $\lambda_N$, $\lambda_Z$ and $\lambda_l$ are the Lagrange multipliers corresponding to constraints on the neutron number $N$, proton number $Z$, and mass multipole moment $q_{l0}$, respectively. The corresponding constrained equations are
\begin{eqnarray}
\langle\psi|\hat{Z}|\psi\rangle&=&Z,\label{HFB2}\\
\langle\psi|\hat{N}|\psi\rangle&=&N,\\
\langle\psi|\hat{Q}_{l0}|\psi\rangle&=&q_{l0}\label{HFB4},
\end{eqnarray}
with $\hat{Q}_{l0}$ defined as
\begin{eqnarray}
\hat{Q}_{l0}&\equiv&(1+\delta_{l,2})\sqrt{\frac{4\pi}{2l+1}}\sum_{i=1}^{A}r_i^lY_{l0}(\theta_i,\phi_i).
\end{eqnarray}
The HFB energy of the constrained system $|\psi_{(N,Z,\{q_{l0}\})}\rangle$ is
\begin{eqnarray}
E^{HFB}_{(N,Z,\{q_{l0}\})}&=&\langle\psi_{(N,Z,\{q_{l0}\})}|\hat{H}|\psi_{(N,Z,\{q_{l0}\})}\rangle.
\end{eqnarray}
The HFB equations (Eqs.~(\ref{HFB1}) to (\ref{HFB4})) are solved iteratively by expanding the single-particle wave-functions onto an axial two-center harmonic oscillator basis. The basis parameters have been automatically optimized for each set of constraints. The Lagrange multipliers are adjusted at each iteration in order that the constraint conditions (Eqs.~(\ref{HFB2}) to (\ref{HFB4})) are fulfilled, and the generalized density matrix is corrected accordingly. This technique allows us to solve the constrained HFB equations in the restricted variational space orthogonal to the PES for which the constraint conditions are fulfilled. In this way, the problems of non-uniform mesh and inaccessible regions mentioned in e.g. \cite{Staszczak:2010} are completely avoided. For more information on the adjustment of the Lagrange multipliers in the case of multiple linear constraints, see for example Appendix~A of~\cite{Younes:2009}.

\subsection{Example of a discontinuity}

\begin{figure}[t]
 \begin{center}
  \includegraphics[width=9cm]{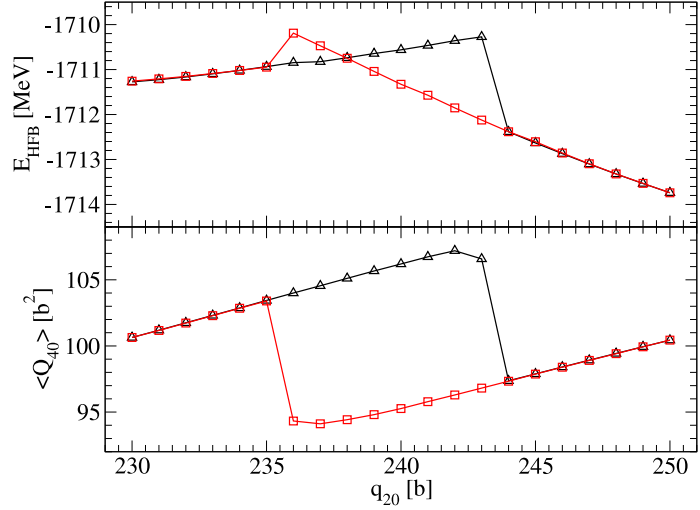}
 \end{center}
 \caption{\label{fig1} (Color online) One-dimensional SPES for $^{226}$Th nucleus with a $q_{30}=52\textrm{ b}^{3/2}$ constraint. Propagation of the calculation is from the left neighbor (black triangles) or from the right neighbor (red squares). Upper panel: total binding energy; lower panel: mean value of the hexadecapole operator $\hat{Q}_{40}$.}
\end{figure}

During the production of a self-consistent potential energy surface, a strange behavior can sometimes be observed. For a given set of constraints, starting the iterative process from two different points can lead to two very different solutions, both converged. This is clearly not what is expected from a self-consistent method, which is supposed to always converge to the only solution minimizing the total binding energy under the action of the constraints, whatever the starting point. Such a strange behavior is shown on Fig. \ref{fig1}, where two one-dimensional SPES have been produced, each point being the starting point for its left (squares) or right (triangles) neighbor. One can see that for a rather large range of constraint values, two different solutions can exist. The ``bow tie'' shape formed by these two curves is characteristic of a discontinuity in the SPES.

\begin{figure}[t]
 \begin{center}
  \includegraphics[angle=-90,width=13cm]{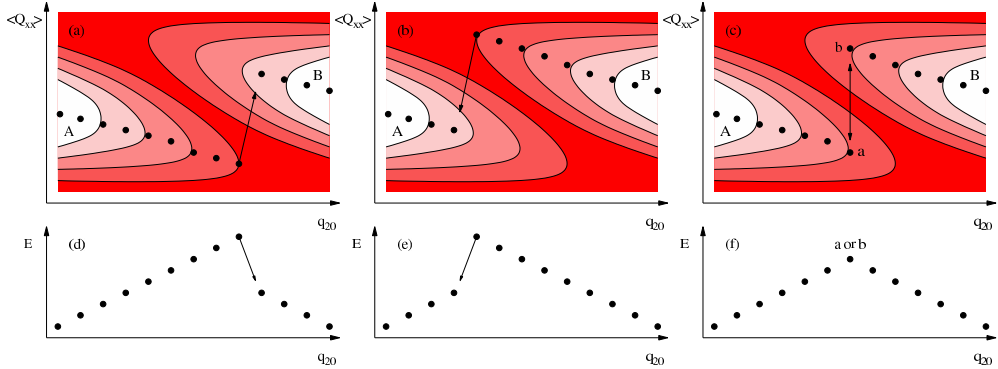}
 \end{center}
 \caption{\label{fig2} (Color online) Illustration of the solutions found when the propagation is from the left neighbor (panels (a) and (d)), from the right neighbor (panels (b) and (e)), and in the ideal case of a perfect minimization (panels (c) and (f)). The upper panels show the same schematic two-dimensional SPES as contour lines, with two valleys labeled A and B. The lower panels show the total binding energy as a function of the $Q_{20}$ values. In panel (f), the point with the maximum total binding energy can be the point $a$ or $b$ from panel (c).}
\end{figure}

\subsection{Explanation}

For an hypothetical system, let us consider the two-dimensional SPES plotted as contour lines on the upper panels (a), (b) and (c) of Fig. \ref{fig2}, showing two valleys A and B in the ($\hat{Q}_{20}$, $\hat{Q}_{xx}$) subspace. Let us now try to produce the one-dimensional SPES with increasing values of $q_{20}$, i.e. each point being the starting point for the calculation of its right neighbor. The resulting total binding energy is plotted on the panel (d), and the mean value $\langle Q_{xx} \rangle$ is plotted as dots on the panel (a). One can see that the converged solutions follow valley A until the system pass the barrier and ``falls'' into valley B. If we do the same calculations with decreasing values of $q_{20}$, we obtain the curves on panels (b) and (e). This time, the system stays longer in valley B before falling into valley A. If we superpose the two curves on the panels (d) and (e), we see the same ``bow tie'' shape as in Fig. \ref{fig1}.

The typical hysteresis figure shown on the lower panel of Fig.~\ref{fig1} or in panels (a) and (b) of Fig.~\ref{fig2} is due to the fact that the system can stay in its valley of origin even when the solution is no longer the global minimum. This phenomenon is mainly related to the way the minimization is performed. In the ideal case of a hypothetical perfect minimization, the global minimum is always found, leading to curves in the panels (c) and (f).

To solve the hysteresis problem in the case of a behavior like the one shown in Fig. \ref{fig1}, one has to discard the non-global minima, keeping the states with the lowest binding energies. This solution leads us to the previous ideal case: every constrained HFB solution is an absolute minimum, there is no more hysteresis, thus no more influence of the starting points. In the following, we will assume that the minimization is perfect, or that hysteresis problems have been tracked and solved. However, even with a perfect minimization, the passage from one valley to the other (from point $a$ to point $b$) is not continuous, since the mean value $\langle Q_{xx} \rangle$ is quite different for these two points. This discontinuity is not visible in the total binding energy curve, since point $a$ and point $b$ can be found to have very close (if not the same) total binding energies. A discontinuity in a SPES can have serious consequences.

\subsection{Consequences}

We define a path in a subspace as a 1-dimensional set of contiguous points in this subspace. In the case of a non-regular mesh, the contiguousness condition becomes a closest neighbors condition (cf. \ref{delaunay} for a definition of this condition). If a path in the constrained subspace crosses a discontinuity, there exists a subspace in which its set of points is not a path anymore, since the points before and after the discontinuity belong to different valleys and are no more contiguous. A description of the evolution of a system along a path that crosses a discontinuity misses the description of the passage between these valleys. This missing passage may correspond to a high-energy saddle point.

One can generalize this statement to the case of $N$-dimensional SPES: any description of the evolution of a system on a discontinuous SPES is an incomplete description, since when crossing a discontinuity, the real system does not ``teleport'' itself from one valley to the other.

An other consequence is that the energy barrier between valleys A and B is necessarily wrongly evaluated, since there can be a huge energy barrier or a low-energy path between both valleys. The barrier height that can be seen on the panel (f) of Fig.~\ref{fig2} may be totally different from the real barrier, and there is no way to estimate this difference by only considering the evolution of the total binding energy like on Fig.~\ref{fig1} or on the panel (f) of Fig.~\ref{fig2}. In other words, the visible saddle point $a$ or $b$ on the panel (f) is a {\it fake} saddle point. If the evolution of the energy on the panel (f) were to be monotonic, a saddle point would not even be visible, and point $a$ or $b$ would correspond to a {\it missing} saddle point.

\section{Finding the discontinuities}

In this section we propose a method to estimate if a given SPES is continuous or not, and to localize and identify the discontinuities.

\subsection{Definitions}

What is usually called a self-consistent potential energy surface (SPES) is a set of $N$ self-consistent solutions $\{|\psi_i\rangle\}_{i=1,N}$ under the action of $N_c$ constraint operators $\{\hat{Q}_j\}_{j=1,N_c}$ with constraint values $q_{i,j}$.
This set can be regularly or randomly distributed on the constrained subspace.
We suppose that a boundary of this SPES is given, separating the deformation subspace in an interior $I$ containing the SPES and an exterior. For example, a SPES with constraint values on $\hat{Q}_{20}$ regularly spaced from 100 b to 200 b with a step $q_{20}^{s}=10$ b has an interior defined as $I=[100,200]$.

Let us introduce the idealized self-consistent potential energy surface (ISPES): for a given SPES, the unique corresponding ISPES is the infinite set of self-consistent solutions $\{|\psi'_x\rangle\}$ under the action of the same $N_c$ constraint operators $\{\hat{Q}_j\}_{j=1,N_c}$ with constraint values $q_{x,j}$ taking all possible values in the same interior. The $x$ quantity is a $N_c$-dimensional vector of coordinates. The ISPES corresponding to the SPES in the previous example can be obtained by taking the limit $q_{20}^{s}\rightarrow 0$:

\begin{eqnarray}
\{|\psi'_x\rangle\}_{x\in I}&=&\lim_{q_{20}^{s}\rightarrow 0} \left(\{|\psi_i\rangle\}\right)\label{plop1}.
\end{eqnarray}

We define the continuity of an ISPES: an ISPES is continuous if the mean values of every one-body multipole operator applied to the system are continuous with respect to each constraint coordinate:

\begin{eqnarray}
&&\forall \hat{Q}_{\lambda\mu}, \forall x\in I,\forall j \in\{1,N_c\},\nonumber\\
&&\lim_{d \rightarrow 0} \langle \psi'(q_{x,j}+d)|\hat{Q}_{\lambda\mu}|\psi'(q_{x,j}+d)\rangle\nonumber\\
&&=\langle \psi'(q_{x,j})|\hat{Q}_{\lambda\mu}|\psi'(q_{x,j})\rangle\label{plop2}
\end{eqnarray}
For the ISPES of the previous example, the continuity condition is
\begin{eqnarray}
&&\forall \hat{Q}_{\lambda\mu}, \forall x\in I,\nonumber\\
&&\lim_{d \rightarrow 0} \langle \psi'(q_{20}(x)+d)|\hat{Q}_{\lambda\mu}|\psi'(q_{20}(x)+d)\rangle\nonumber\\
&&=\langle \psi'(q_{20}(x))|\hat{Q}_{\lambda\mu}|\psi'(q_{20}(x))\rangle\label{plop3}
\end{eqnarray}

We then define the continuity of a SPES: a SPES is continuous if its corresponding ISPES is continuous.

If a SPES is not continuous, one can identify a couple of neighboring states that surround a discontinuity of the mean-value of one of the possible operators on the corresponding ISPES, thus localizing this discontinuity on the SPES.

For a given discontinuity, at least one of the mean-values of the usual multipole moments must be discontinuous. We call the lowest-order discontinuous multipole moment operator the {\it signature of the discontinuity}.

\subsection{Density distance}

As a way to numerically estimate the difference between two neighboring states $|\psi\rangle$ and $|\psi'\rangle$, we define the density distance $D_{\rho\rho'}$ as
\begin{equation}
D_{\rho\rho'}\equiv\int d\vec{r}\textrm{ }|\rho(\vec{r})-\rho'(\vec{r})|
\end{equation}
with $\rho(\vec{r})$ and $\rho'(\vec{r})$ being the local spatial density of the $|\psi\rangle$ and $|\psi'\rangle$ states, respectively. This distance is dimensionless, and can be seen as the geometrical difference between two states, measured in nucleons. Any state of $A$ nucleons has a density distance of $A$ with the void state ($\int d\tau^3 \rho(\vec{r})=A$ and $\rho'(\vec{r})=0$).

\subsection{Maximum density distance}

\begin{figure}[t]
 \begin{center}
  \includegraphics[width=7cm]{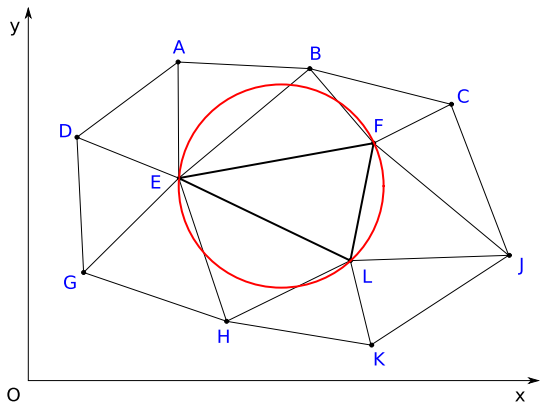}
 \end{center}
 \caption{\label{figdelaunay} Example of a Delaunay triangulation in a 2-dimensional space, using the usual Euclidean distance. Points $E$, $F$ and $L$ belong to the same Delaunay cell. In a $N$-dimensional space, a set of $(N+1)$ points is a Delaunay cell if its circumscribed circle/sphere/\ldots contains no other point. The circumscribed circle of the Delaunay cell $EFL$ is shown in red.}
\end{figure}

We define the maximum density distance $D_{\textrm{max}}$ of a SPES as
\begin{eqnarray}
&&D_{\textrm{max}}\equiv \textrm{max}(D_{\rho_i\rho_j})\nonumber\\
&&\textrm{with }i\in\{1,N\},j\in \mathcal{N}_i,
\end{eqnarray}
$\mathcal{N}_i$ being the set of closest neighbors of states $|\psi\rangle_i$. In the case of a regular mesh, two points are closest neighbors if only one of their coordinates differ and if this difference is minimal. In the case of a non-regular mesh, two points are considered as closest neighbors if there exists a Delaunay cell containing them~\cite{Barber:1996}\label{delaunay}. Being given a $N_c$-dimensional set of points and an Euclidean distance function, the Delaunay triangulation is unique (if no points are co-circular). The distance function used for the Delaunay cells computation has to be carefully chosen in order to take into account variations in all $N_c$ possible directions. For example, to find the closest neighbors in the case of a non-regular 4-dimensional mesh with constraints on ($\hat{Q}_{20}$,$\hat{Q}_{30}$,$\hat{Q}_{40}$ and $\hat{Q}_{60}$), we have used the following pseudo-Euclidean distance function:
\begin{eqnarray}
d^2&\equiv& \left(\frac{q_{20}-q'_{20}}{1 \textrm{b}}\right)^2+\left(\frac{q_{30}-q'_{30}}{1 \textrm{b}^{3/2}}\right)^2\nonumber\\
 &+&\left(\frac{q_{40}-q'_{40}}{1 \textrm{b}^2}\right)^2+\left(\frac{q_{60}-q'_{60}}{1 \textrm{b}^3}\right)^2
\end{eqnarray}
with the values of $q_{20}$, $q_{30}$, $q_{40}$ and $q_{60}$ being given in b, b$^{3/2}$, b$^2$ and b$^3$, respectively. A 2-dimensional example of a Delaunay triangulation using the usual Euclidean distance is given on Fig.~\ref{figdelaunay}.

\begin{figure}[t]
 \begin{center}
  \includegraphics[width=9cm]{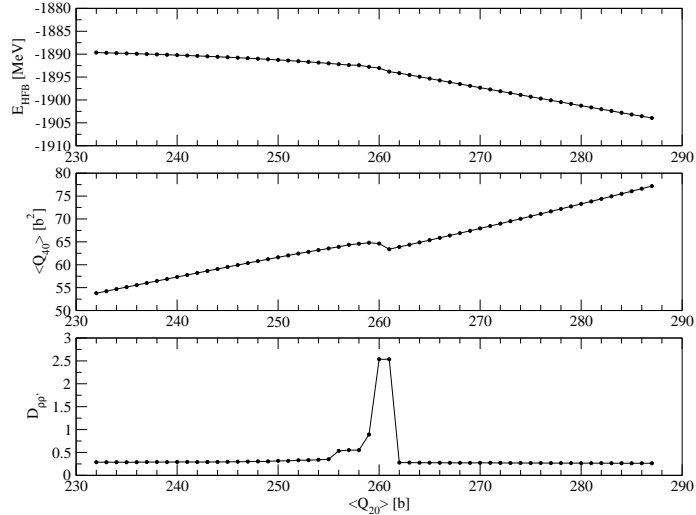}
 \end{center}
 \caption{\label{fig3} Example of a discontinuity for the symmetric scission of $^{256}$Fm. The upper panel shows the total binding energy, the center panel shows the mean value of the hexadecapole operator $\hat{Q}_{40}$, and the lower panel shows the maximum total density distance between each point and its closest neighbors. The cylindrical symmetry is conserved.}
\end{figure}

\begin{figure}[t]
 \begin{center}
  \includegraphics[width=9cm]{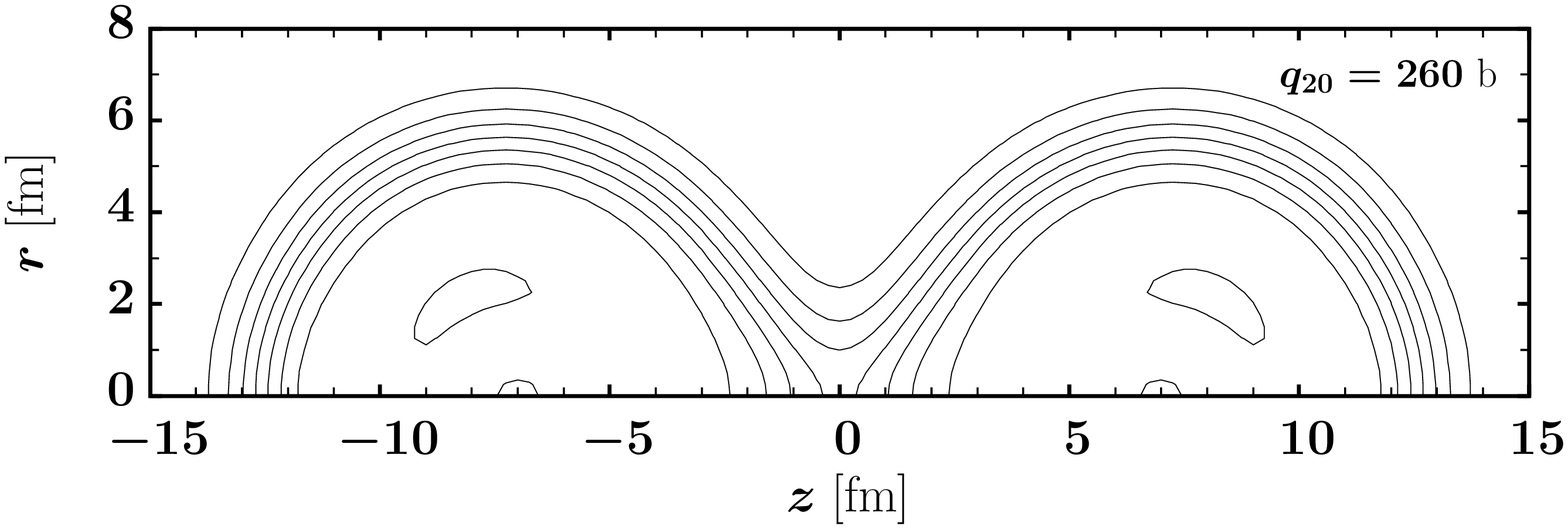}
  \includegraphics[width=9cm]{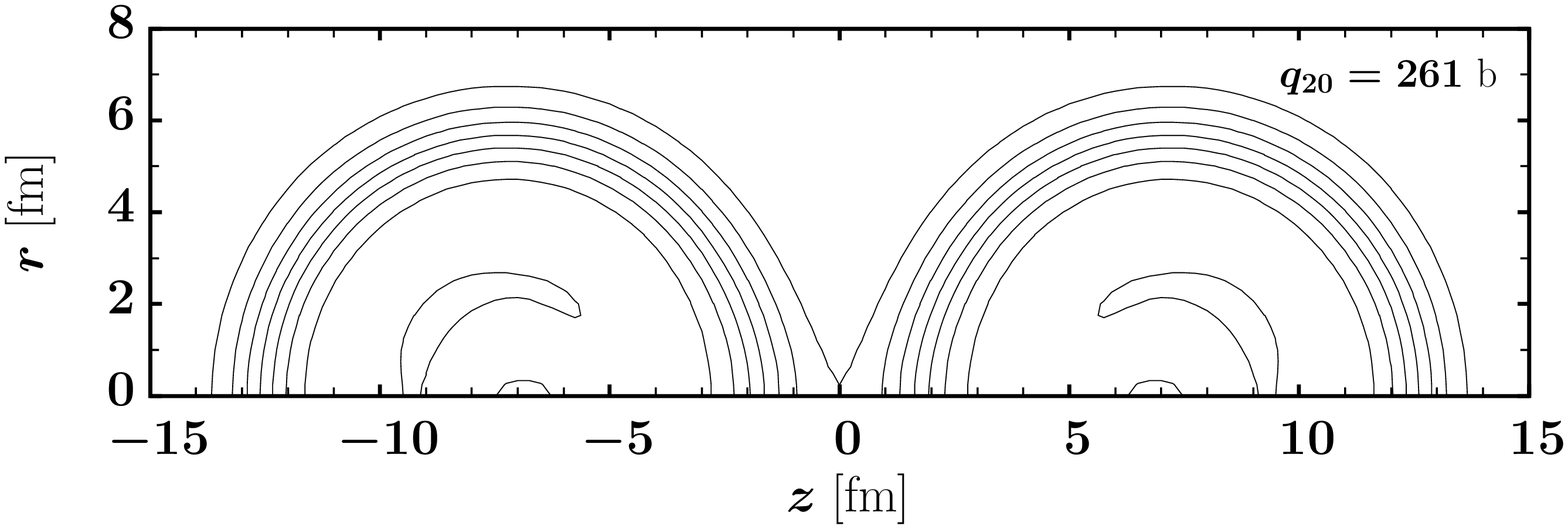}
 \end{center}
 \caption{\label{fig4}Local densities for $^{256}$Fm in the ($z$, $r$) plane before (top) and after (bottom) the discontinuity shown on Fig.~\ref{fig3}. The isolines are separated by 0.02~fm$^{-3}$.}
\end{figure}

An example of a discontinuity for the symmetric fission of $^{256}$Fm is shown on Fig.~\ref{fig3}. The evolution of the HFB energy is rather smooth (top panel of Fig.\ref{fig3}). There is no way to suspect a possible discontinuity just by considering the HFB energy curve. However, a spike of the $D_{\rho\rho'}$ value (bottom panel of Fig.~\ref{fig3}) is observed for $q_{20}=260$ b and $q_{20}=261$ b, consequence of a discontinuity. We can see that the mean value of the hexadecapole operator $\hat{Q}_{40}$ changes at this discontinuity (middle panel of Fig.~\ref{fig3}), so we can conclude that this discontinuity has $\hat{Q}_{40}$ for signature. Since the evolution of the energy is monotone, no saddle point is visible, but there has to be a saddle point between both valleys in the $\hat{Q}_{40}$ subspace. This is an example of a {\it missing} saddle point. The local densities before and after the discontinuity are shown on Fig.~\ref{fig4}. The main visible change in the local densities seems to be a stronger neck between the nascent fission fragments.

\begin{figure}[t]
 \begin{center}
  \includegraphics[width=9cm]{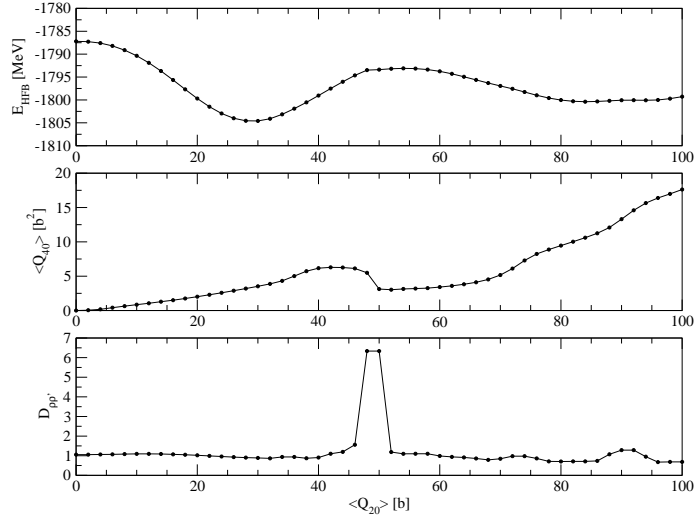}
 \end{center}
 \caption{\label{fig5} Example of a discontinuity for the symmetric first barrier of $^{240}$Pu. The upper panel shows the total binding energy, the center panel shows the mean value of the hexadecapole operator $\hat{Q}_{40}$, and the lower panel shows the maximum total density distance between each point and its closest neighbors. The cylindrical symmetry is conserved.}
\end{figure}

\begin{figure}[t]
 \begin{center}
  \includegraphics[width=9cm]{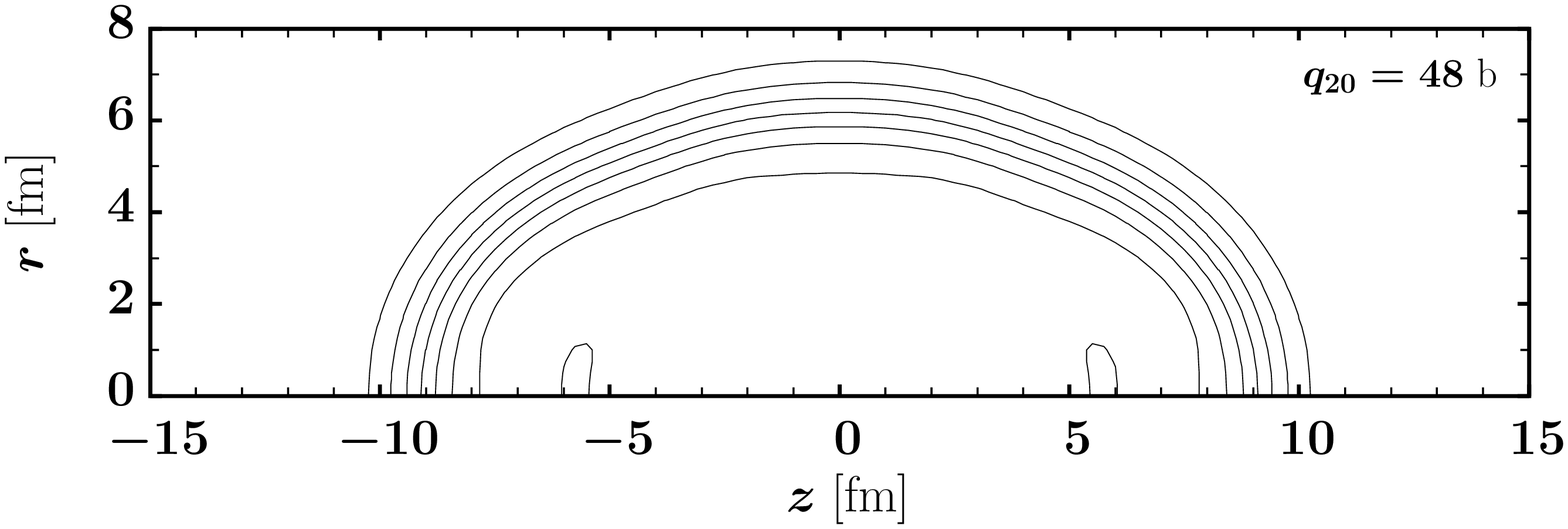}
  \includegraphics[width=9cm]{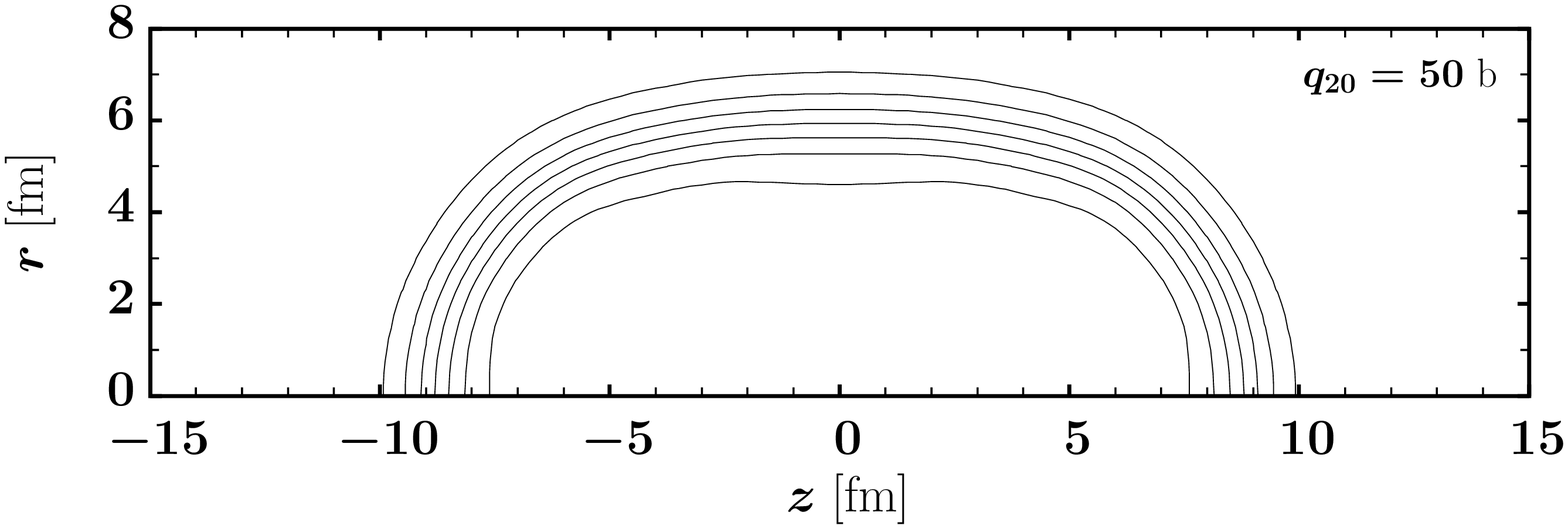}
 \end{center}
 \caption{\label{fig6}Local densities for $^{240}$Pu in the ($z$, $r$) plane before (top) and after (bottom) the discontinuity shown on Fig.~\ref{fig5}. The isolines are separated by 0.02~fm$^{-3}$.}
\end{figure}

An other example of a discontinuity is shown on Fig.~\ref{fig5}, for the symmetric first barrier of ${}^{240}$Pu.
We would like to stress that this axial $1$-dimensional SPES is shown as an example only, and must not be considered as a prediction of the real first barrier of $^{240}$Pu.
The influence of the triaxial degree of freedom is neglected here, and it has been shown that this degree of freedom significantly lowers the first barrier of several actinides \cite{Abusara:2010,Girod:1983,Pashkevich:1969,Moller:1970,Rutz:1995,Egido:2000,Bonneau:2004,Lu:2011}.
The situation is very similar to the one on Fig.~\ref{fig3}: the evolution of the HFB energy is smooth (top panel of Fig.~\ref{fig5}), but a spike of the $D_{\rho\rho'}$ value (bottom panel of Fig.~\ref{fig5}) indicates that there is a discontinuity between $q_{20}=48$ b and $q_{20}=50$ b.
The signature of this discontinuity is the hexadecapole operator $\hat{Q}_{40}$, since its mean value is discontinuous as shown in the middle panel of Fig.~\ref{fig5}.
The local densities corresponding to the states before and after the discontinuity are shown on Fig.~\ref{fig6}.
The visible difference between these states seems to be the quantity of matter around the $z=0$ plane.
Since the discontinuity occurs at a saddle point, this point is what we have called a {\it fake} saddle point in the Introduction.

Since the step for the $q_{20}$ constraint value is smaller for the SPES of Fig.~\ref{fig3} compared to the SPES of Fig.~\ref{fig5}, it is important to remark that the average value of $D_{\rho\rho'}$ for continuous points is smaller (around 0.3 for $^{256}$Fm versus 1.0 for $^{240}$Pu).
To check the continuity of a SPES, the used steps must be small enough to be able to see spikes corresponding to discontinuities.


\section{Obtaining a continuous SPES}

As far as we know, all of the discontinuities we have had to deal with led at least to a density distance of several nucleons. It is then safe to consider that a SPES is continuous if we have $D_{\textrm{max}}<2$. In order to reach this level, all of the discontinuities have to be ``cleaned''. There are two ways to do so.

\subsection{Additional dimension}

\begin{figure}[t]
 \begin{center}
  \includegraphics[width=11.2cm]{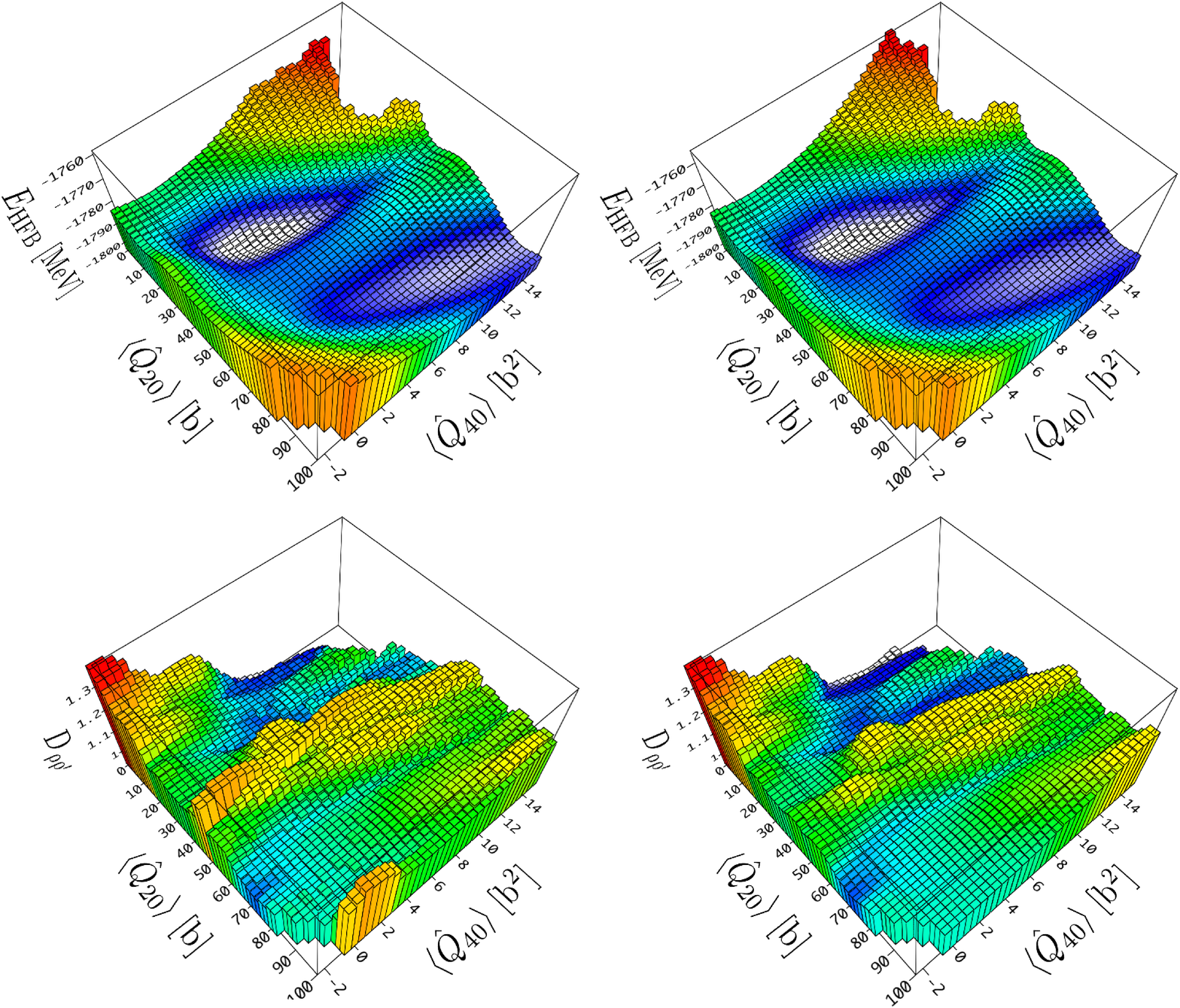}
 \end{center}
 \caption{\label{fig7} (Color online) Self-consistent potential energy surfaces for the symmetric first barrier of $^{240}$Pu in the ($q_{20}$, $q_{40}$) subspace. The top figures show the evolution of the total binding energy, the lower figures show the corresponding maximum total density distance between each point and its closest neighbors. For the left figures, an automatic optimization of the basis parameters has been used for each point, while the basis parameters have been kept constant for the right figures.}
\end{figure}

The first way to remove a discontinuity is simply to globally extend the mesh with the signature of this discontinuity. For example, if a given one-dimensional SPES along $\hat{Q}_{20}$ has a discontinuity with $\hat{Q}_{40}$ for signature, the two-dimensional SPES in the ($\hat{Q}_{20}$, $\hat{Q}_{40}$) subspace will not have any discontinuity with $\hat{Q}_{40}$ for signature. This method is brute-force, and can lead to a high number of points.
As an example, we have extended the SPES shown on Fig.~\ref{fig5} by taking the signature of the discontinuity $\hat{Q}_{40}$ as a constraint. The resulting SPES is shown on Fig.~\ref{fig7}.
On this figure, the topology of the two valleys (corresponding to the first and second well) is very similar to the one of the schematic valleys shown on Fig.~\ref{fig2}.
One can see that the automatic basis optimization can be a source of rather small local maxima of $D_{\rho\rho'}$ values, since these maxima values are lowered when disabling the basis optimization. Anyway, these local maxima are small enough to let us consider that the SPES shown on Fig.~\ref{fig7} are continuous.

\subsection{Connecting points}

The second way to remove a discontinuity is to locally extend the mesh with some missing points at a discontinuity. For the schematic discontinuity of Fig.~\ref{fig2}, one could simply add to the SPES the connecting points between valleys A and B of the panel (f). To find these connecting points, one calculates a small two-dimensional SPES large enough to include points $a$ and $b$ like the one plotted in panel (c), one finds the saddle point between the valleys by calculating the discrete local derivatives for each point, or by a more subtle technique if needed, like for example the {\it immersion technique}~\cite{Moller:2009,Luc:1991}, and one calculates the least-energy paths from this saddle point to the bottom of each of the valleys. The {\it immersion technique} is a fast and reliable method to find the saddle point between two given points on a regular mesh. Its principle is to mimic the way a liquid would fill a valley before flowing into an other one through a saddle point.

\begin{figure}[t]
 \begin{center}
  \includegraphics[angle=-90,width=6cm]{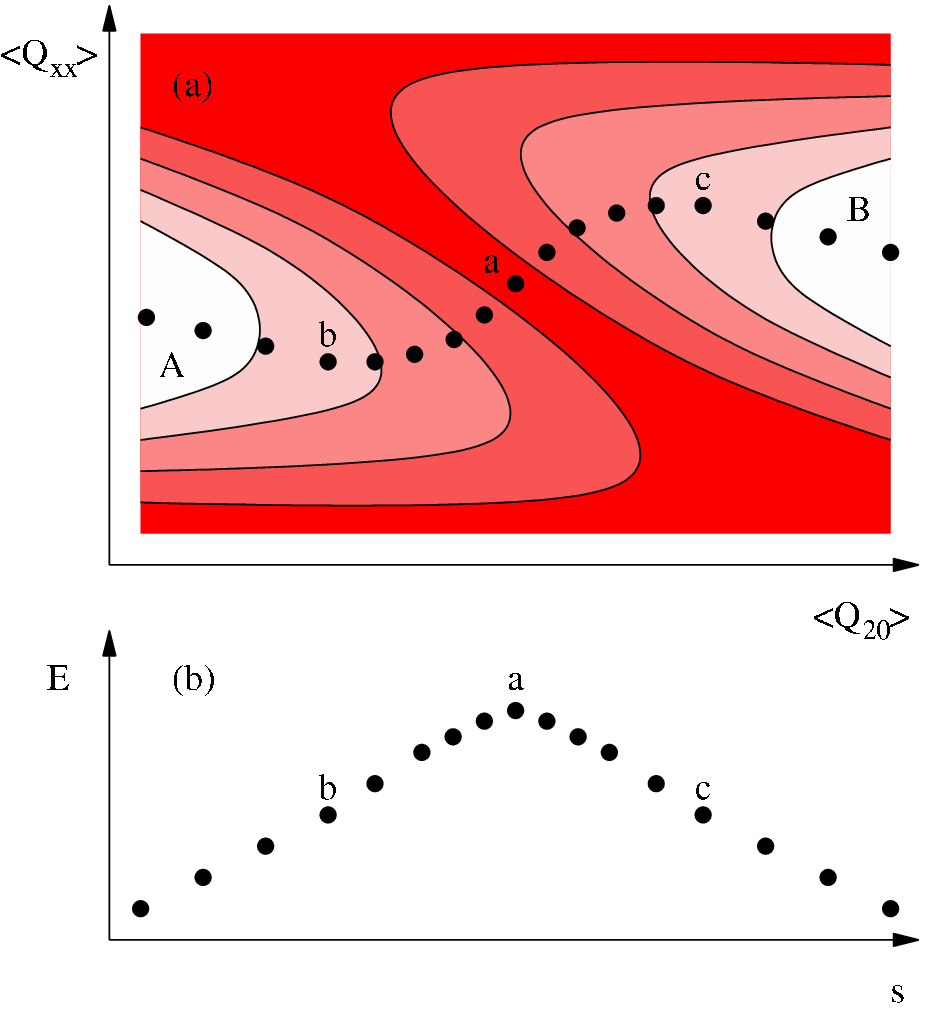}
 \end{center}
 \caption{\label{fig8} Illustration of the ``connecting points'' method. The saddle point $a$ is found, then points $a$ to $b$ and $a$ to $c$ are found as least action paths. Original points before $b$, original points after $c$, and both least action paths ($a$ to $b$ and $a$ to $c$) are then joined to form a continuous path. The constraint value $q_{20}$ from Fig.~\ref{fig2} has been replaced with a curvilinear abscissa $s$ in lower panel.}
\end{figure}

The schematic result is shown on Fig.~\ref{fig8}. It is worth mentioning that $\langle \hat{Q}_{20}\rangle$ is no longer a good variable for a one-dimensional description, for two reasons. Firstly, it is possible that the least-energy path ``backbends'', i.e. there can be multiple points with the same $\langle \hat{Q}_{20}\rangle$ value. Secondly, the connecting points are not anymore minimizing the total binding energy for a given $\langle \hat{Q}_{20}\rangle$ value. This is why a curvilinear abscissa is used instead of $\langle \hat{Q}_{20}\rangle$, and why the original points between $b$ and $c$ are discarded and replaced with points from the least action paths. The resulting set of points can no longer be called a SPES, but can be called an effective PES.

This method is rather convenient in the case of one-dimensional SPES, but can be hard to automatize for SPES of higher dimension. The fact that the constraint variables are no longer good variables can also be problematic when trying to perform dynamical propagations on the resulting effective PES.

\subsection{Discussion}

We would like to stress that for a given system, the dimension of the smallest continuous SPES is the dimension of the smallest corresponding continuous ISPES. If the continuous description of a physical process needs $N_p$ degrees of freedom, the dimension of a corresponding continuous SPES must obey $N\ge N_p$. Trying to obtain a continuous description with an insufficient number of degrees of freedom with a change of variables like in the ``connecting points'' method presented above can be dangerous for the physical interpretation of the resulting quantities like the inertia tensor, mainly because the fact that some points do not minimize the total binding energy anymore for a given $\langle \hat{Q}_{20}\rangle$ value forbids the use of this latter as a good variable.

Using the ``additional dimension'' method presented above does not lead to such problems, but can be numerically difficult. For some physical processes, having $N=N_p$ can mean that the number of points of the continuous SPES is huge if the considered mesh is rectangular and regular. As a way to lower this number of points, one can restrict the SPES to the solutions with a low energy (non-rectangular mesh), and/or use multidimensional adaptive meshing (non-regular mesh). 

\section*{Conclusion}

Discontinuities are not easily visible, and have important consequences. Any result extracted from a discontinuous SPES should be considered with extreme caution. The method to obtain $D_{\textrm{max}}$ that we have presented in this paper is simple, and allows to check with good confidence if a given SPES is continuous or not, without any additional self-consistent calculation. The value of $D_{\textrm{max}}$ should be specified for any SPES, like the conserved symmetries and the level of convergence reached.

In the near future, we plan to perform dynamical calculations on SPES for the description of the fission process and the prediction of some fission fragment properties, using a Time-Dependent Generator Coordinate Method~\cite{Goutte:2005,Dubray:2008}, and the continuity of the produced SPES will be carefully checked using the method presented here. One of the main advantages of this method is that it can be fully automatized, allowing an on-the-fly detection of discontinuities during the production of a SPES.

In this study, the definition of $D_{\rho\rho'}$ has been chosen for practical reasons (mainly its numerical simplicity), but other expressions can be used to estimate the ``distance" between two neighboring states, involving for example the non-local density
\begin{equation}
D'_{\rho\rho'}\equiv\int\int d\vec{r}\textrm{ }d\vec{r}'\textrm{ }|\rho(\vec{r},\vec{r}')-\rho'(\vec{r},\vec{r}')|
\end{equation}
or the overlap between neighboring states
\begin{equation}
D''_{\rho\rho'}\equiv\frac{1}{|\langle \psi|\psi'\rangle|}.
\end{equation}
Such expressions can reveal a discontinuity in non-local fields, but their computational cost may be important. One should also note that the proposed expression does not take into account time-odd quantities. When calculating a SPES for an odd nucleus, another expression should be used if one is looking for discontinuities in time-odd quantities.
In Ref.~\cite{Moller:2009}, the authors state that ``[\ldots] minimization does not necessarily lead to the wrong solution; in many simple problems it will give the correct saddle point. Unfortunately, one cannot know {\it a priori} whether a true or incorrect solution will be found''. Thanks to the simple method presented in this paper to calculate $D_{\textrm{max}}$, it is now easy to estimate if and where a discontinuity has occurred while calculating a SPES. One can consider that a SPES with a low enough $D_{\textrm{max}}$ value does not contain such ``incorrect solutions'' ({\it fake} or {\it missing} saddle points), and that the ``Problem SPES" from Introduction is fixed. By comparison, there is to our knowledge no numerical method to fix ``Problem NSPES" from the introduction, in other words, to estimate if partial local extrema obtained by non-self-consistent methods in a given subspace remain partial local extrema in bigger subspaces.

\section*{Acknowledgments}

We gratefully thank D.~Gogny for his useful advices concerning this study.

\bibliographystyle{elsarticle-num}
\bibliography{dubray}

\end{document}